\documentclass[aps,pra,print,twocolumn]{revtex4} 
\usepackage{amsthm,amsmath,amsfonts,amssymb,amscd}
\usepackage{graphicx}
\usepackage{verbatim}
\usepackage{setspace} 
\usepackage{mathrsfs} 
\usepackage{epstopdf} 

\usepackage{hyperref}
\hypersetup{
    bookmarksnumbered=true, 
    unicode=false, 
    pdfstartview={FitH}, 
    pdftitle={}, 
    pdfauthor={}, 
    pdfsubject={}, 
    pdfcreator={}, 
    pdfproducer={}, 
    pdfkeywords={}, 
    pdfnewwindow=true, 
    colorlinks=true, 
    linkcolor=blue, 
    citecolor=blue, 
    filecolor=blue, 
    urlcolor=blue 
}

\newcommand{\ket}[1]{|#1\rangle}
\newcommand{\bra}[1]{\langle#1|}

\begin{document}

\title{Quantum imaging by coherent enhancement}

\author{Guang Hao Low, Theodore J. Yoder, and Isaac L. Chuang}
\affiliation{Center for Ultracold Atoms, Research Laboratory of Electronics, and Department of Physics,
Massachusetts Institute of Technology, Cambridge, Massachusetts 02139, USA}
\date{\today}

\begin{abstract}
Conventional wisdom dictates that to image the position of fluorescent atoms or molecules, one should
stimulate as much emission and collect as many photons as possible. That is, in this classical case, it has
always been assumed that the coherence time of the system should be made short, and that the statistical
scaling $\sim1/\sqrt{t}$ defines the resolution limit for imaging time $t$. However, here we show in
contrast that given the same resources, a long coherence time permits a higher resolution image. In this
quantum regime, we give a procedure for determining the position of a single two-level system, and
demonstrate that the standard errors of our position estimates scale at the Heisenberg limit as $\sim 1/t$,
a quadratic, and notably optimal, improvement over the classical case.
\end{abstract}

\pacs{03.67.-a, 42.50.-p, 06.30.Bp}

\maketitle
The precise imaging of the location of one or more point objects is a problem ubiquitous in science and
technology. While the resolution of an image is typically defined through the diffraction limit as the
wavelength $\sim\lambda$ of illuminating light, the final estimate of object position instead exhibits a
shot-noise limited precision $\sigma$ that scales with the number of scattered photons detected -- a
consequence of the law of large numbers. Thus, in the absence of environmental noise, it is the time allowed
for accumulating statistics that appears to limit the precision of position measurements.

Surprisingly, when the objects to be imaged are imbued with quantum properties, these well-known classical
limits on resolution and precision can be improved. Impressive sub-optical resolutions of $\sim
\frac{\lambda}{10}$~\cite{Betzig1992,Hell2007} are obtainable by advanced microscopy~\cite{Hell2007} protocols
such as STED~\cite{Trifonov2013}, RESOLFT~\cite{Hofmann2005}, STORM~\cite{Rust2006}, and
PALM~\cite{Betzig2006}.
Each in its own way exploits the coherence of a quantum object by storing its position $x_i$ in its quantum
state $\ket{\psi}$ over an extended period of time. Ultimately however, even for state-of-art, it is still the
statistical scaling $\sigma\sim\frac{1}{\sqrt{t}}$ that limits the precision of a position estimate taking
time $t$.

Yet, fundamentally, coherent quantum objects allow for a precision scaling quadratically better, as
$\sigma\sim\frac{1}{t}$. This so-called Heisenberg limit~\cite{Demkowicz2012} is a fundamental restriction of
nature that bounds the precision of a single-shot phase estimate of $\ket{\psi}$, i.e. given a single copy of
$\ket{\psi}$, to $\sim\frac{1}{t}$, a bound attainable in the regime of long
coherence~\cite{Nielsen2004,Berry2009,Higgins2009}.

\begin{figure}
\includegraphics[width=1.0\columnwidth]{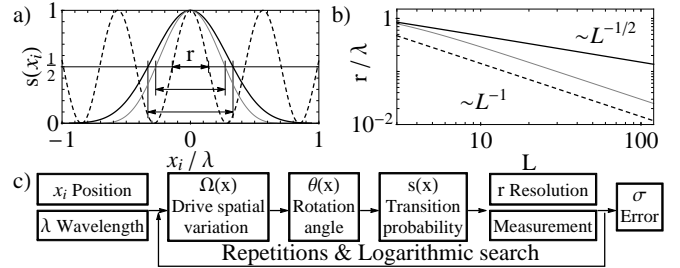}
\caption{\label{OptimalUM}
a) Map from position $x_i$ to transition probability $s(x_i)$. This is ideally unambiguous with a single
narrow peak of width $r$ (solid). The ambiguous map has multiple peaks (dashed). b) Scaling of $r$ with the
number of coherent drive pulses $L$. The optimal scaling is $\sim1/L$ (dashed, thin), but often suboptimal
$\sim1/\sqrt{L}$ for unambiguous maps (thick).
c) Procedure outline for estimating $x_i$ with error $\sigma$ scaling at the Heisenberg limit. This combines
an optimal $r$-scaling unambiguous map with measurement in a logarithmic search.
}
\end{figure}

How then can quantum coherence be fully exploited to minimize the time required to obtain an estimate of a
quantum object's position $x_i$ with standard error $\sigma$?
An apparent contradiction arises since photon scattering rates approach zero in the limit of infinite
coherence, in contrast to traditional imaging, where maximizing scattering is desirable.
A similar problem arises in magnetic resonance imaging, but is there resolved by a two-step
process: map $x_i$ coherently to $\ket{\psi}$,
and then read out $\ket{\psi}$ using just a few photons.
However, current approaches have two flaws. First, the mapping is
typically ambiguous (Fig.~\ref{OptimalUM}a). Due to the periodicity of quantum phases, multiple $x_i$ can be
encoded
into the same observable of $\ket{\psi}$ -- often the transition probability $s(x_i)$.
Second, the mapping resolution $r$ -- the length scale over which $s(x_i)$ varies -- cannot be improved
arbitrarily in an effective manner. Doing so, with say a long sequence of $L$ coherent excitations, either
introduces more ambiguity or requires time that does not perform better than the statistical scaling
(Fig.~\ref{OptimalUM}b). Approaches that estimate position with Heisenberg-limited scaling must overcome these
two challenges.

Such well-known difficulties are apparent when using a spatially varying coherent drive, e.g. a gaussian beam,
that produces excitations varying over space $\sim\lambda$.
Due to projection noise~\cite{Itano1993}, $s(x_i)$ can only be estimated with error scaling
$\sim\frac{1}{\sqrt{t}}$. Thus for any given $r$, a precision $\sigma\sim\frac{r}{\sqrt{t}}$ results.
Working around projection noise and improving these resolutions is the focus
of much work in magnetic resonance as well quantum information science with trapped
ions~\cite{Wineland1998,Vitanov2011,Shappert2013,Shen2013,LeSage2013,Merrill2014}. Unfortunately,
state-of-art~\cite{Vitanov2011,Jones2013B,Low2014} excitation sequences, or pulse sequences, that produce a
single unambiguous peak are sub-optimal -- they offer a resolution of $r\sim\frac{\lambda}{\sqrt{L}}$
(Fig.~\ref{OptimalUM}) which is no better than the statistical scaling.

We present a new procedure that images quantum objects with
precision $\sigma \sim \frac{1}{t}$, using a two-step imaging process
which unambiguously maps spatial position to quantum state, allowing
for readout with imaging resolution that scales as the optimum
achievable by the Heisenberg limit.
Like prior art, a pulse sequence is employed to implement the unambiguous mapping. In contrast though, we
develop new sequences with the optimum resolution scaling $r\sim \frac{1}{L}$ (Fig.~\ref{OptimalUM}). Due to
the narrowness of $r$, measuring the quantum state is much more likely to tell one where the object is not,
rather than where it is located. Thus, our optimal unambiguous mapping alone is insufficient for achieving
$\sigma \sim\frac{1}{t}$.
However, this issue is neatly resolved using a logarithmic search,
modeled after quantum phase estimation~\cite{Nielsen2004,Berry2009,Higgins2009}, that applies our mapping
several times with varying widths.
This logical flow (Fig.~\ref{OptimalUM}c) leads to our final result: an imaging algorithm with optimal
precision $\sigma \sim \frac{1}{t}$.
From the classical perspective that imaging should be done with short coherence times and maximal photon
scattering, our algorithm is a complete surprise. In fact, our results imply that the best method for imaging
quantum objects is to collect very few photons from a source that can be coherently controlled.

We begin by defining the resources required for imaging the position of a quantum object in one dimension. The
action of pulse sequences on this system is briefly reviewed to demonstrate the mapping of spatial position to
transition probability. This framework allows us to define the unambuguity and optimality criteria for a
transition probability. We show that our new pulse sequences have both properties. These same properties also
enable an efficient logarithmic search for system position, solving the projection noise issue. We then
discuss estimates of real-world performance, generalizations to higher dimensions and multiple objects.

Consider a quantum two-level system with state $\ket{\psi}\in \text{SU}(2)$ at an unknown position $x_i\in I$
contained in a known interval $I$. Measurements in the $\{\ket{0},\ket{1}\}$ basis are assumed, for
simplicity.  Provided is a coherent drive, over which we have phase $\phi$ and duration $t$ control, with a
known spatially varying Rabi frequency $\Omega(x)$, where $x=x_i-x_c$ can be translated by arbitrary distance
$x_c$. With this coherent drive, a unitary rotation $U_{\phi}[\theta]=e^{-i\frac{\theta}{2}
[\cos{(\phi)}\hat{X}+\sin{(\phi)}\hat{Y}]}$, where $\hat{X},\hat{Y}$ are Pauli matrices, that traverses angle
$\theta(x)=\Omega(x)\tau$ can be applied to $\ket{\psi}$. Chaining $L$ such discrete rotations generates a
pulse sequence $\mathcal{S}=U_{\phi_L}[\theta]...U_{\phi_1}[\theta]\equiv(\phi_1,..,\phi_L)$. When applied to
$\ket{0}$, this results in the state $\mathcal{S}\ket{0}$ and the transition probability $p(\theta)=|\langle
1|\mathcal{S}|0\rangle|^{2}$ in $\theta$ coordinates. As $\theta$ depends on position $x_i$, a map from
spatial coordinates to transition probability is achieved through $s(x)=p(\theta(x))$.

\begin{figure}
\includegraphics[width=1.0\columnwidth]{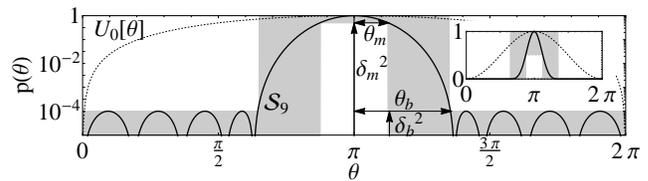}
\caption{\label{ComparisonPlot} Transition probability $p(\theta)$ of the sequence $\mathcal{S}_L$ (solid)
plotted for $L=9$ in comparison to a single rotation $U_0[\theta]$ (dotted). The range of the envelope
$p(\theta;\theta_b,\theta_{m})$ is shaded. Primary features of $\mathcal{S}_L$ are sidelobes of uniform
bounded error $\delta_b^2$, and a central peak with width parameters $\theta_b,\theta_{m}$ that scale as
$\sim1/L$. The inset plots the same on a linear scale.}
\end{figure}

The criteria of unambiguity and optimality can now be expressed as constraints on the form of $s(x)$.
Unambiguity means that $s(x)$ has only a single sharp peak in some domain of $x$ so that excitation with high
probability only occurs if the system lies in some small contiguous region of space $\Delta$, which defines,
through its width $|\Delta|$, the resolution $r\approx|\Delta|$.  Noting that $p(\theta)$ is periodic in
$\theta\rightarrow \theta\pm2\pi$ and, for odd $L$, necessarily peaks at $p(\theta=\pi)=1$, unambiguity is
possible if no other large peaks occur in the domain of $0\le\theta(x)<2\pi$ and $\theta(x)$ varies
monotonically with $x$. For example, a Gaussian diffraction-limited beam with spatial profile
$\Omega(x)=\Omega_0 e^{-x^2/4\lambda^2}$ restricted to $x>0$ and the choice $\Omega_0 t=\sqrt{e} \pi$ suffices
and will be used in what follows. With this choice, and assuming unambiguous $p(\theta)$, the only peak in
$s(x)$ occurs at $x=\sqrt{2}\lambda$, exactly where $\theta=\pi$, and where also the gradient of $\Omega(x)$
is steepest so that the width of the peak is minimized. Expressed in $\theta$ coordinates, the peak width of
$2\theta_b=|\Delta|\theta'$ is illustrated in Fig.~\ref{ComparisonPlot}, where
$\theta'=\max_{x}\big|\frac{d\theta(x)}{dx}\big|$. In particular, optimality means that the width of this
single peak scale like $\theta_b\sim\frac{1}{L}$ -- any better scaling would permit a means to beat the
Heisenberg limit.

Both unambiguity and optimality are satisfied by our new family of pulse sequences $\mathcal{S}_L$, which
realize the transition profile
\begin{align}
p_L{(\theta;\delta_b)}&=\left|\frac{T_L\left[\beta_L(\delta_b)\sin{(\frac{\theta}{2})}\right]}{T_L\left[\beta_L(\delta_b)\right]}\right|^{2},
\end{align}
plotted in Fig.~\ref{ComparisonPlot}, where $T_L[x]=\cos{\left[L \cos^{-1}{(x)}\right]}$ is the
$L^{\text{th}}$ Chebyshev polynomial, and $\beta_L(\delta_b)=T_{L^{-1}}\left[\delta_b^{-1}\right]$. Primary
features of the function $p_{L}(\theta;\delta_b)$ include an optimally narrow, like $\sim\frac{1}{L}$, central
peak given a uniform bound $\delta_b^2$ on sidelobes~\cite{Dolph1946}. We find it useful to consider the
half-width $\theta_b$ of the central peak at sidelobe height $\delta_b^2$ and $\theta_m$ at arbitrary heights
$\delta_m^2>\delta_b^2$ (Fig.~\ref{ComparisonPlot}) with ratio of widths $R=\theta_b/\theta_m>1$:
\begin{align}
\label{Eq:widths}
 \theta_b(L)&=\frac{2}{L}\text{arcsech}(\delta_b)+\mathcal{O}\left(\frac{1}{L^3}\right),
 \\\nonumber
R&=1\Big/\sqrt{1-\left(\frac{\text{arcsech}(\delta_b/\delta_m)}{\text{arcsech}(\delta_b)}\right)^{2}}+\mathcal{O}\left(\frac{1}{L^2}\right),
\end{align}

The phases $(\phi_1,..,\phi_L)$ that implement $\mathcal{S}_L$ for arbitrarily large $L$ are elegantly
described in closed-form. We first consider the broadband variant
$\mathcal{S}_L^{\text{B}}=(\chi_1,...,\chi_L)$ which realizes
$p_L^\text{B}(\theta;\delta_b)=1-p_L(\theta-\pi;\delta_b)$ and is related to $\mathcal{S}_L$ via the toggling
transformation $\phi_k=(-1)^k\chi_k+2\sum_{h=1}^{k-1}(-1)^h\chi_h$~\cite{Levitt1986}. That the three pulse
member $\mathcal{S}_3^{\text{B}}=(\chi,0,\chi)$ has
$\chi=2\tan^{-1}{\Big[\tan{\left(\pi/3\right)}\sqrt{1-\beta_3^{-2}(\delta_b)}\Big]}$ is easily verified. As
the phases of $\mathcal{S}_3^{\text{B}}$ form a palindrome~\cite{Jones2013B,Low2014}, $\mathcal{S}_3^{B}$
implements an effective rotation of angle $\theta_e$, defined through
$1-p_L^B(\theta;\delta_b)=\cos^{2}{(\theta_e/2)}$, about some axis in the $\hat{x}$-$\hat{y}$ plane. Thus
replacing each base pulse in $\mathcal{S}_{L_2=3}^{\text{B}}[\delta_b]$ with a different sequence
$\mathcal{S}_{L_1=3}^{\text{B}}[1/\beta_{L_{2}}(\delta_b)]$ produces the transition profile $p_{L_1
L_2}^\text{B}(\theta;\delta_b)$ by repeatedly applying the semigroup property $T_{n}[T_{m}[x]]=T_{nm}[x]$ of
Chebyshev polynomials. For $L_2=3$ and any odd $L_1$, this corresponds exactly to the transition profile of
$\mathcal{S}_{3 L_1}^{\text{B}}[\delta_b]=(\psi,0,\psi)\circ
\mathcal{S}_{L_1}^{\text{B}}[1/\beta_{3}(\delta_b)]$, where $\circ$ defines a nesting operator
$(a_1,a_2,...)\circ(b_1,b_2,...)=(a_1+b_1,a_1+b_2,...,a_2+b_1,a_2+b_2,...)$. As we provide $L_1=3$, by
induction the phases required for $\mathcal{S}_{3^n}^{\text{B}}[\delta_b]$ and $\mathcal{S}_{3^n}[\delta_b]$
can be obtained in closed form as a function of $\delta_b$ for all $n\in \mathbb{Z}^{+}$.

After $\mathcal{S}_L$ is applied for some choice of beam position $x_c$, a measurement of $\ket{\psi}$
extracts encoded positional information.
As visualized with the envelope in Fig.~\ref{ComparisonPlot}:
\begin{align}
\label{promise_filter}
p(\theta;\theta_b,\theta_m)=\begin{cases}
\ge \delta_{m}^2, & |\theta-\pi|\le\theta_m, \\
\in[0,1],&\theta_m<|\theta-\pi|<\theta_b,\\
\le \delta_b^2, & \text{otherwise},
\end{cases}
\end{align}
if $\ket{1}$ is obtained after a measurement, the object is located with high probability in the central peak,
corresponding to the spatial interval $\Delta^b$
of width $|\Delta^b|=2\theta_b/\theta'$ centered on $x_c$. Conversely, if $\ket{0}$ is obtained, then the
object is located outside, in
$I\backslash \Delta^m$, with high probability, where $\Delta^m$ is also centered on $x_c$ with width
$|\Delta^m|=2\theta_m/\theta'$.
However, projection noise means that false positives or negatives can still occur. Fortunately, these can be
made exponentially improbable by
initializing to $\ket{0}$, and taking $l$ repeats.

The probability $P$ of an incorrect classification, that is, assigning an estimate $x_e$ to an interval that
does not contain $x_i$ is an elementary exercise in probabilities. We summarize: Over $l$ repetitions, we
measure the outcome $\ket{1}$ $k$ times. If $k/l\ge \bar{p}=(\delta_m^2+\delta_b^2)/2$, we assign
$x_e\in\Delta^b$. Else, we assign $x_e\in I\backslash \Delta^m$. Thus
\begin{align}
\label{Eq.ErrorProbability}
P&=\text{max}(P_1,P_2)\le\exp{\left[-l(\delta_m^2-\delta_b^2)^2/2\right]},\\ \nonumber
P_1&=\text{Pr}[x_e\in
I\backslash\Delta^m|x_i\in\Delta^m]=\text{Pr}\left[k/l<\bar{p}\middle|x_i\in\Delta^m\right],\\ \nonumber
P_2&=\text{Pr}[x_e\in \Delta^b|x_i\in I\backslash \Delta^b]=\text{Pr}\left[k/l\ge\bar{p}\middle|x_i\in
I\backslash \Delta^b\right],
\end{align}
where $P$ is bounded by Hoeffding's inequality applied to binomial distributions~\cite{Hoeffding1963}. Thus
$x_e$ can be reliably classified to either inside or outside a region of width $|\Delta|\sim\frac{1}{L}$ in a
constant number of $\sim 1$ measurements.

\begin{figure}
\includegraphics[width=1.0\columnwidth]{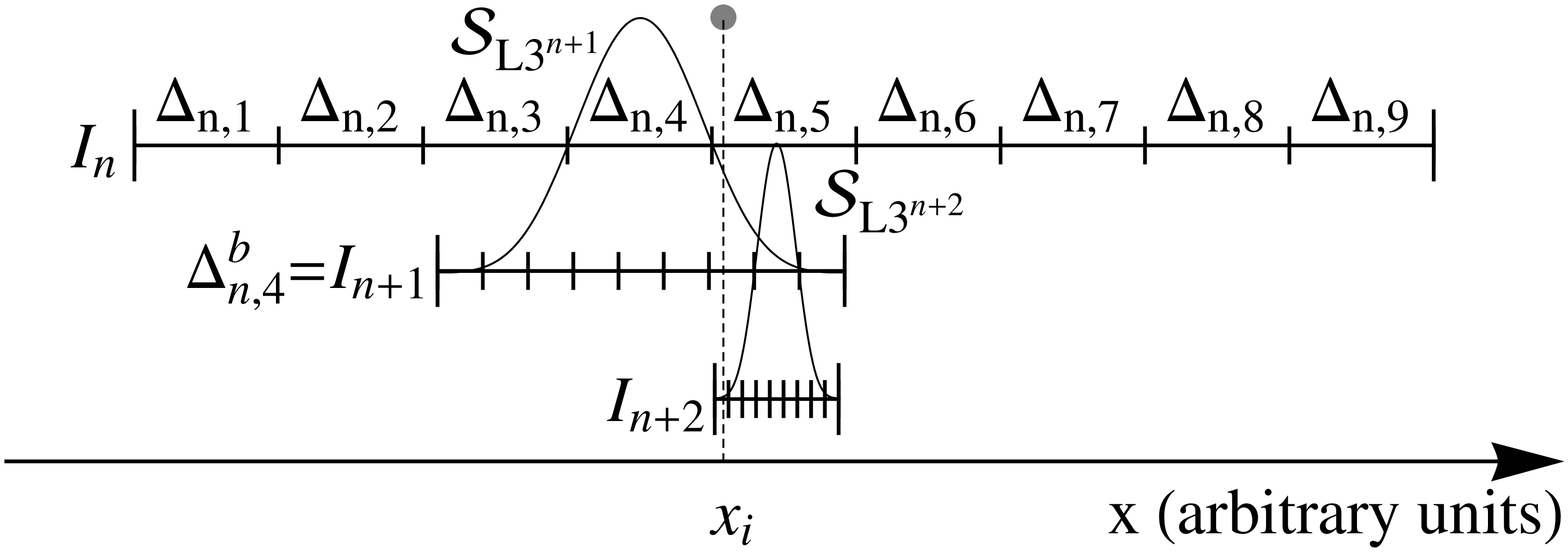}
\caption{\label{Fig:Subdivision}Two iterations of the logarithmic search illustrated for an object located at
$x_i$. At the $n^{\text{th}}$ iteration, the object has been localized to the interval $x_i\in I_n$. $I_n$ is
split into $D$ intervals $\Delta_d$, and $\mathcal{S}_{L3^{n+1}}$ with base width $\Delta_d^b$ is applied to
each interval $l$ times. The first positive classification to $\Delta_d^b$ for any $d$ further narrows the
object position to $x_i \in \Delta_b^d = I_{n+1}$. In this example, $R|\Delta_d^b|\approx|I_n|$, $R\approx3$,
$\delta_m^2=1/2$, $\delta_b^2=10^{-4}$, $K=3$.}
\end{figure}

A key insight allows us to sidestep the $\sigma\sim\frac{1}{\sqrt{t}}$ scaling of projection noise arising
from accumulating statistics indefinitely. Once the object has been classified to some interval $\Delta^b$
with high probability, a sequence that is $K$ times longer than $\mathcal{S}_{L}$ can be applied to query
subintervals of width $K$ times smaller than $\Delta^b$. As the width of these subintervals scale optimally
like $\sim\frac{1}{L}$ in Eq.~\ref{Eq:widths}, it is never profitable, in the coherent regime, to accumulate
more than a constant number of statistics. Rather, $L$ should be increased in geometric progression as far as
coherence times allow. In other words, imaging proceeds by logarithmic search, illustrated in
Fig.~\ref{Fig:Subdivision} for $x_i\in I_0$ initially known only to be in the region $I_0$ of width
$|I_0|\lessapprox \lambda$. Although this process is conceptually similar to binary search, we must account
for two key differences: 1) queries are corrupted by projection noise and 2) the accept and reject intervals
are asymmetric i.e. $\Delta^b\neq\Delta^m$.

The search is initialized by choosing the largest $L_0$ such that
$|I_0|<2\theta_{b}(L_0)/\theta'$
followed by $n=1,...,M$ iterations of a recursive process. The $n^{\text{th}}$ iteration involves three steps.
First, $I_{n-1}$ is split into $D$ smaller subintervals of equal width, each centered on $x_{n,d}$. Here,
$D=\lceil K R \rceil$, $K\in \mathbb{Z}^+$, and $d=1,...,D$.
Second, the classification procedure involving $l$ applications of $\mathcal{S}_{L_n}$, where $L_n=L_0 K^n$,
is then applied for each $d$ with $x_c = x_{n,d}$ until for some $d$, $x_{e}$ is classified into
$x_{e}\in\Delta^b_{n,d}$. Third, we update
$I_{n}=\Delta^b_{n,d}$, which is of width $|I_{n}|= \frac{|I_{n-1}|}{K}$. By induction over $M$ iterations,
$x_e$ lies in an interval width $|I_M|\approx|I_0|\left(\frac{1}{K}\right)^M$. Since $K>1$, exponential
precision $|I_M|$ is achieved in only a linear number of $\mathcal{O}(M)$ state initializations and
measurements! Any misclassification of $x_e \in \Delta^b_{n,d}$ such that $x_i \notin \Delta^b_{n,d}$ will be
detected in the next iteration as the probability of misclassifying $x_{e}\in\Delta^b_{n+1,d}$ again becomes
vanishingly small like $\mathcal{O}(P^2)$, as seen from Eq.~\ref{Eq.ErrorProbability}. In that case, the
previous iteration is repeated.
Assuming $x_i\in I_M$ is uniformly distributed, the standard deviation is
$\sigma\approx\frac{|I_{0}|}{K^M\sqrt{12}}\big(1+\mathcal{O}(P)\big)$.

The runtime $ t $ of this logarithmic search is a geometric sum over iterations $n=1,...,M$, each involving an
expected number $E=Dl/2+\mathcal{O}(P)$ applications of $\mathcal{S}_{L_n}$. Letting, $\Omega'=\theta'/\tau$,
we have
\begin{align}
\label{Eq:TimeCost}
t&= E \sum_{n=1}^{M}\tau L_n
\approx  \frac{E|I_0|\theta'L_0K}{K-1}\frac{K^M-1}{K^M}\frac{1}{|I_M|\Omega'} \\ \nonumber
 &\approx \frac{4EK \text{arcsech}(\delta_b)}{K-1}\frac{1}{|I_M|\Omega'}
 \approx \frac{2EK \text{arcsech}(\delta_b)}{\sqrt{3}(K-1)}\frac{1}{\sigma\Omega'},
\end{align}
where the approximations $K^M-1\approx K^M$, $|I_0|\approx 2\theta_b(L_0)/\theta'$, $\sigma\approx
|I_{M}|/\sqrt{12}$ are made.

Thus, in Eq.~\eqref{Eq:TimeCost} we have arrived at our final result: an estimate of object position $x_e$
with standard deviation $\sigma=\frac{1}{\Omega'}\mathcal{O}(\frac{1}{t})$ exhibiting a Heisenberg-limited
scaling with time, and requiring $M=\mathcal{O}(\log{\frac{1}{\sigma}})$ measurements.
The most straightforward minimization of the constant factors requires the choices of 1) shortest wavelength
$\lambda$ and 2) strongest drive $\Omega_0$. However, these parameters are often fixed by experimental
constraints. One could then optimize over the independent variables $\delta_b,\delta_m,l,K$. For example,
inserting $K=3, l=5, \delta_b^2=\frac{7}{20},\delta_m^2=\frac{13}{20}$ into Eq.~\ref{Eq:TimeCost}, evaluating
$E$ to $\mathcal{O}(P^2)$, and Eq.~\ref{Eq.ErrorProbability} exactly gives $t\approx\frac{26}{\Omega'
\sigma}$.

\begin{figure}
\includegraphics[width=1.0\columnwidth]{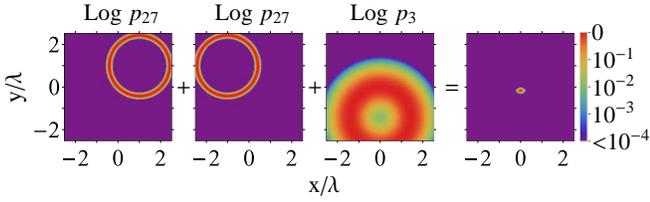}
\caption{\label{Fig:MaximumLikelihoodCircles}(Color online) Demonstration of using three Gaussian beams in two
dimensions, $\theta(x,y)=\sqrt{e}\pi e^{-(x^2+y^2)/4\lambda^2}$, and exploiting the strongly peaked log
transition probability $\log{p_L}\equiv\log{p_L(\theta(x,y);10^{-2})}$ to locate the object at
$(x_i,y_i)=(0,0)$. The beam centers are placed at ${(1,1),(-1,1),(0,-\sqrt{2})}$ and sequences
$\mathcal{S}_{27},\mathcal{S}_{27},\mathcal{S}_3$ implemented, respectively.}
\end{figure}

Notably, our imaging procedure only gracefully degrades in the presence noise found in real systems with any
finite coherence time $\tau_c$. Noise replaces $\mathcal{S}$ with an implementation-dependent quantum channel
$\mathcal{E}(\rho)$ acting on the initial state $\rho_i=\ket{0}\bra{0}$ to produce the output
$\rho_{noise}=\mathcal{E}(\rho_i)$, in comparison to the ideal case of
$\rho_{ideal}=\mathcal{S}\rho_i\mathcal{S}^{\dagger}$. As the trace distance~\cite{Nielsen2004}
$\text{TrD}(\rho_{ideal},\rho_{noise})=\gamma$ bounds the difference in measurement probabilities using any
measurement basis, noise shifts the envelope in Eq.~\ref{promise_filter} by $\delta_b^2\rightarrow
\delta_b^2+\gamma,\delta_m^2\rightarrow\delta_m^2-\gamma$ and modifies the misclassification probability in
Eq.~\ref{Eq.ErrorProbability} to
\begin{align}
\label{Eq.HoeffingsNoise}
P\le \exp{\left[-l(\delta_m^2-\delta_b^2-2\gamma)^2/2\right]}\ll 1,
\end{align}
As long as $\gamma <\frac{1}{2}$, classification succeeds \emph{independent} of the noise model as we can
always satisfy Eq.~\ref{Eq.HoeffingsNoise} by some choice of $\delta_{b}$, $\delta_m$, and
$l(\gamma)\propto(\delta_m^2-\delta_b^2-2\gamma)^{-2}$. Success for $\gamma\ge\frac{1}{2}$ depends on details
of the nose model. Of course, $\gamma$ generally increases with sequence length, such as in a completely
depolarizing channel where $\gamma_n=\frac{1}{2}(1-e^{-t L_n/\tau_c})$. For fixed $\delta_m,\delta_b$, the
runtime in Eq.~\ref{Eq:TimeCost} becomes $t\propto\sum_{n=1}^M l(\gamma_n) K^n.$ As the final precision
$\sigma\propto\frac{1}{K^M}$, the instantaneous scaling in the presence of noise
\begin{align}
\frac{d t}{d(\sigma^{-1})}&=\frac{d t}{d M}\frac{d M}{d(\sigma^{-1})}\propto
\frac{1}{(\delta_m^2-\delta_b^2-2\gamma_{M})^2}\\ \nonumber
&\propto 1+\frac{2}{\delta_m^2-\delta_b^2}\frac{\tau L_M}{\tau_c}+\mathcal{O}\left((\tau L_M/\tau_c)^2\right)
\end{align}
shows clearly a continuous degradation from the noiseless Heisenberg-limited scaling
$\lim_{\tau_c\rightarrow\infty}\frac{d t}{d(\sigma^{-1})}\propto1$ to the statistical scaling $\frac{d
t}{d(\sigma^{-1})}\propto \sqrt{t}$.
In the regime of strong decoherence at $\tau L_M\sim \tau_c$ where higher orders dominate, accumulating
statistics with $\mathcal{S}_{L_M}$ and applying the law of large numbers becomes more time-efficient than
using logarithmic search and correcting many misclassifications.

Generalizing our imaging scheme to higher dimensions is straightforward. Finding the $(x_i,y_i,z_i)$
coordinates of a object in three dimensions is reducible to three separate one-dimensional problems by using
three cylindrical Gaussian beams oriented about orthogonal axes with spatial profiles
$\theta(x,y,z)=\sqrt{e}\pi e^{-s^2/4\lambda^2},\,s\in\{x,y,z\}$. Illustrated in
Fig.~\ref{Fig:MaximumLikelihoodCircles} is how one can also use three Gaussian beams with radial symmetry
$\theta(x,y)=\sqrt{e}\pi e^{-(x^2+y^2)/4\lambda^2}$ to query the object position in two dimensions.

Extending the procedure to multiple $Q>1$ objects also presents no fundamental difficulty. If there are $Q$
objects, during iteration $n$, the classification procedure should be applied to \emph{all} subintervals
$\Delta^m_{n,d}$. Then, all subintervals that return a positive classification, i.e. $x_e\in\Delta^b_{n,d}$
are subject to subdivision and classification in the $(n+1)^{\text{th}}$ iteration. In particular, crosstalk
can be suppressed by decreasing $\delta_b^2$ by factor $\sim Q$. Therefore, in time $t\sim \frac{Q}{\sigma}$,
all subintervals that contain objects will be found.

Many avenues of further inquiry are facilitated by the optimally narrow pulse sequences applied here for
imaging. For example, the functional form of these pulse sequences match Dolph-Chebyshev window
functions~\cite{Dolph1946,Harris1978} which have been studied in the context digital signal
filtering~\cite{Fettweis1984,Fettweis1986}. This hints at a deeper connection where the extensive machinery
developed for signal processing could be applied to pulse sequences, interpreted as \emph{quantum}
filters~\cite{Sorae2014}.
Additionally, while the language of optical regimes of operation has been used here, the techniques presented
are extremely generic and apply to the entire electromagnetic spectrum. With a fidelity of $\sim 10^{-6}$ per
rotation in a pulse sequence, object positions can in principle be estimated with precision $\sim
10^{-6}\lambda$ in time scaling at the Heisenberg limit. At optical wavelengths, a practical limit may be
imposed by the finite size of atoms, but exciting possibilities include using microwave wavelengths of $\sim
1$cm to measure nanoscale $\sim 10$nm features, or using radio waves in high-resolution magnetic resonance
imaging, where instead of using magnetic field background gradients to provide nuclei or quantum dots
spatially dependent resonance conditions, the spatial varying amplitude of the radio-frequency drive itself is
used in conjunction with nuclear spins, which are known to have extremely long coherence times.

GHL acknowledges support from the ARO Quantum Algorithms
program. TJY acknowledges support from the NSF iQuISE IGERT program. ILC acknowledges support from the NSF
CUA.

\end{document}